\pdfoutput=1
\documentclass[11pt,a4paper,draft=false]{report}
\usepackage{amsmath,amsthm,amssymb}
\usepackage{bm,txfonts}       
\usepackage{breqn}              
\usepackage{cancel}
\usepackage{siunitx}            
\usepackage{arydshln}         
\usepackage{booktabs}         
\usepackage{color,soul}         
\usepackage[inline]{enumitem}   
\usepackage{fullpage}
\usepackage[pdftex]{graphicx}
\DeclareGraphicsExtensions{.pdf,.png,.jpg}
\usepackage{caption}
\usepackage[caption=false,labelfont=bf]{subfig}
\usepackage{listings}
\usepackage{url}
\usepackage[pdfpagelabels,pdfusetitle,colorlinks=true,pdfborder={0 0
  0},pdfauthor={Dimitar Stanev},colorlinks=false]{hyperref}
\usepackage{natbib}
\usepackage[acronym]{glossaries}

\newacronym{dofs}{DoFs}{Degrees of Freedom}
\newacronym{dof}{DoF}{Degree of Freedom}
\newacronym{eoms}{EoMs}{Equations of Motion}
\newacronym{fd}{FD}{Forward Dynamics}
\newacronym{pd}{PD}{Proportional Derivative}
\newacronym{cns}{CNS}{Central Nervous System}
\newacronym{lr}{LR}{Lateral Rectus}
\newacronym{sr}{SR}{Superior Rectus}
\newacronym{ir}{IR}{Inferior Rectus}
\newacronym{mr}{MR}{Medial Rectus}
\newacronym{so}{SO}{Superior Oblique}
\newacronym{io}{IO}{Inferior Oblique}
\newacronym{em}{EOMs}{Extraocular Muscles}
\newacronym{fl}{F-L}{Force-Length}
\newacronym{fv}{F-V}{Force-Velocity}
\newacronym{fpe}{F-PE}{Passive-Force-Length}
\newacronym{ft}{F-T}{Tendon-Force-Length}
\makeglossaries{}

\DeclareMathAlphabet\mathbfcal{OMS}{cmsy}{b}{n} 

\renewcommand*{\vec}[1]{\bm{#1}}
\newcommand{\R}[1]{\mathfrak{R}^{#1}}
\newcommand{\inr}[1]{\in\R{#1}}

\makeatletter
\@addtoreset{algorithm}{chapter}
\makeatother
\def\equationautorefname~#1\null{Equation #1\null}

\usepackage{makecell}

\title{An Open-Source OpenSim Oculomotor Model for Kinematics and Dynamics
  Simulation}

\author{Konstantinos Filip, Dimitar Stanev\footnote{Electrical and Computer
    Engineering Department, University of Patras, Greece; Corresponding
    author: \url{stanev@ece.upatras.gr}}, and Konstantinos Moustakas}

\date{\today}

\begin{document}

\maketitle

\begin{abstract}
  Physics-based modeling and dynamic simulation of human eye movements has
  significant implications for improving our understanding of the oculomotor
  system and treating various visuomotor disorders. We introduce an open-source
  biomechanical model of the human eye that can be used for kinematics and
  dynamics analysis. This model is based on the passive pulley hypothesis,
  constructed based on the data reported in literature regarding physiological
  measurements of the human eye and made publicly available\footnote{SimTK
    project: \url{https://simtk.org/projects/eye}}. The model is implemented in
  \texttt{OpenSim}, which is an open-source framework for modeling and
  simulation of musculoskeletal systems. The model incorporates an eye globe,
  orbital suspension tissues and six extraocular muscles. The excitation and
  activation patterns for a variety of targets can be calculated using the
  proposed closed-loop fixation controller that drives the model to perform
  saccadic movements in a forward dynamics manner. The controller minimizes the
  error between the desired saccadic trajectory and the predicted
  movement. Consequently, this model enables the investigation muscle activation
  patterns during static fixation and analyze the dynamics of eye movements.
\end{abstract}

\section*{Introduction}\label{sec:introduction}

Rapid and accurate eye movements are of great importance for natural vision and
thus studying human eye movement can improve our understanding of the oculomotor
system and treating various visuomotor disorders~\cite{Lee2006, Wei2010a}. Over
the past decades, biomechanics simulation has provided the means to analyze
different human movements~\cite{Delp2007}. The same principles can be used to
analyze visual tasks by modeling the musculoskeletal properties of the
oculomotor system. Consequently, this model can be used to investigate muscle
activation patterns during static fixation, analyze the dynamics of various eye
movements, calculate metabolic costs and simulate eye disorders, such as
different forms of strabismus~\cite{Wong2004}. Furthermore, it can be easily
integrated with available full body models in order to analyze the relation
between the vestibular and oculomotor systems.

Eye movements are a generated from the coordinated activation of the six
\gls{em}. Clinical trials have provided a profound knowledge on the properties
of the \gls{em} and their line of action on the eye globe~\cite{Robinson1969a}
and the resistive tension of the surrounding tissues~\cite{Collins1981,
  Iskander2018}. Various computational models of the extraocular muscles and
orbital mechanics have been proposed, which provide insight for oculomotor
biomechanics, control of eye movement~\cite{Bach-y-Rita1971} and binocular
misalignment. These models focus on the realism of muscle behavior and they were
based on the viscoelastic properties and physiological data \gls{em}.

The first 3D biomechanical model was developed by~\cite{Robinson1964a,
  Robinson1969}, who simplified the formulation by only considering the
elasticity of the \gls{em} ignoring their dynamics. The model incorporates
anatomically realistic muscle paths and empirical innervation-length-tension
relationships. In order to study the neural control of rapid saccadic movements,
models using anatomical and mechanical properties of \gls{em} have been
developed by accounting for the nonlinear muscle dynamics~\cite{Thelen2003a,
  Millard2013}. Such models, having the advantage of supporting dynamics
simulation, are used in conjunction with brain level
controllers~\cite{Angelaki2004, James2018}.

\section*{Methods}\label{sec:methods}

\subsection*{Eye Modeling}\label{sec:eye-Modeling}

The eye model consists of the eye globe, three pairs of \gls{em} and the
connective passive tissues. The size of an emmetropic human adult eye is
approximately $0.0242 \si{\m}$ (transverse, horizontal), $0.0237 \si{\m}$
(sagittal, vertical), $0.022 - 0.0248 \si{\m}$ (axial, anteroposterior) with no
significant difference between sexes and age groups. In the transverse diameter,
the eye may vary from $0.021 \si{\m}$ to $0.027 \si{\m}$, thus it can be
approximated by a solid sphere of radius $r = 0.012 \si{\m}$. The weight of an
average human eye is $m = 0.0075 \si{\kg}$ and the moment of inertia can be
calculated assuming a spherical homogeneous and isotropic model
$I = 2 / 5 m r^2$. The eye has three rotational \gls{dofs}, namely
incyclotosion-excyclotosion ($\vec{x}$-axis), adduction-abduction
($\vec{y}$-axis) and supraduction-infraduction ($\vec{z}$-axis).

\subsection*{Muscle Modeling}\label{sec:muscle-modeling}

The six \gls{em}, including four rectus muscles and two oblique muscles, are
controlled by the cranial nerves so as to track a visual target and to stabilize
the image of the object of interest on the retina. The \gls{lr} and \gls{mr}
muscles form an agonist/antagonist pair that produce horizontal eye
movements. The \gls{sr} and \gls{ir} muscles form the vertical
agonist/antagonist pair, which mainly controls vertical eye movement and also
affects rotation about the line of sight (secondary action) and the horizontal
plane (tertiary action). The \gls{so} muscle passes through the cartilaginous
trochlea attached to the orbital wall, which reflects the \gls{so} path by
$51 \si{\degree}$. The \gls{io} muscle originates from the orbital wall
anteroinferior to the globe center and inserts on the sclera posterior to the
globe equator. The primary actions of \gls{so} and \gls{io} cause rotation of
the globe around the visual axis, but also affect vertical (secondary action)
and horizontal (tertiary action) movements.

The model relies on the passive pulley assumption, which states that the pulleys
have fixed to the orbit pulley points~\cite{Clark1977,
  Miller2007}. \autoref{tab:eye-muscle-path} shows the positions of the origin,
insertion and pulleys for the \gls{em}, defined in the local body coordinates of
the eye globe. The data are based on physiological
measurements~\cite{Iskander2018}, with some minor modification so as to prevent
unrealistic muscle-surface penetration. Since no position was documented for the
origin of the \gls{so}, a point close to the origins of the rectus muscles was
chosen to match the fiber length in the primary position of the \gls{so} muscle.

\begin{table}[ht]
  \centering
  \caption{Muscle path points for the six \gls{em} defined in the local frame of
    the eye globe (dimensions are given in meters).}\label{tab:eye-muscle-path}
  \resizebox{\textwidth}{!}{
  \begin{tabular}{@{}cccccccccc@{}}
    \toprule
    \textbf{Muscle}
    & \multicolumn{3}{c}{\textbf{Origin}}
    & \multicolumn{3}{c}{\textbf{Pulley}}
    & \multicolumn{3}{c}{\textbf{Insertion}} \\
    \midrule
    & \textit{\textbf{Ox}} & \textit{\textbf{Oy}} & \textit{\textbf{Oz}}
    & \textit{\textbf{Px}} & \textit{\textbf{Py}} & \textit{\textbf{Pz}}
    & \textit{\textbf{Ix}} & \textit{\textbf{Iy}} & \textit{\textbf{Iz}} \\
    \midrule
    \gls{lr} & -0.034 & 0.0006 & -0.013 & -0.0102 & 0.0003 & 0.012 & 0.0065 & 0 & 0.0101 \\
    \gls{mr} & -0.030 & 0.0006 & -0.017 & -0.0053 & 0.00014 & -0.0146 & 0.0088 & 0 & -0.0096 \\
    \gls{sr} & -0.0317 & 0.0036 & -0.016 & -0.0092 & 0.012 & -0.002 & 0.0076 & 0.0104 & 0 \\
    \gls{ir} & -0.0317 & -0.0024 & -0.016 & -0.0042 & -0.0128 & -0.0042 & 0.00805 & -0.0102 & 0 \\
    \gls{so} & 0.0082 & 0.0122 & -0.0152 & -0.030834 & 0.001145 & -0.01644 & 0.0044 & 0.011 & 0.0029 \\
    \gls{io} & 0.0113 & -0.0154 & -0.0111 & -0.00718 & -0.0135 & 0 & -0.008 & 0 & 0.009 \\
    \bottomrule
  \end{tabular}
  }
\end{table}

The Millard muscle model~\cite{Millard2013} has been adopted for the modeling of
the \gls{em}, permitting parameterization of the characteristic curves according
to the experimental measured data. The muscles were modeled using the rigid
tendon assumption that ignores the elasticity of the tendon. This means that the
series element of the muscle model is not included (the tendon length $l^T$ is
equal to the tendon slack length $l_s^T$). \gls{em} are considered
parallel-fibered muscles, so the pennation angle is zero ($\alpha = 0$). The
values for the maximum isometric force $f_o^M$, optimal fiber length $l_o^M$ and
tendon length $l^T$ are presented in~\autoref{tab:eye-muscle-parameters}.

\begin{table}[ht]
  \centering
  \caption{Millard muscle parameters for the
    \gls{em}.}\label{tab:eye-muscle-parameters}
  \begin{tabular}{@{}cccccccccc@{}}
    \toprule
    \thead{Muscle \\ \quad}
    & \thead{Maximum Isometric \\ Force (N)}  
    & \thead{Optimal Fiber \\ Length (m)}
    & \thead{Tendon Slack \\ Length (m)}
    & \thead{Maximum Contraction \\ Velocity (m / s)} \\
    \midrule
    \gls{lr} & 1.4710 & 0.04898 & 0.0084 & 3.8483 \\
    \gls{mr} & 1.5740 & 0.04084 & 0.0038 & 4.6155 \\
    \gls{sr} & 1.1768 & 0.04487 & 0.0054 & 4.2009 \\
    \gls{ir} & 1.4269 & 0.04549 & 0.0048 & 4.1437 \\
    \gls{so} & 0.6031 & 0.03956 & 0.0265 & 4.7648 \\
    \gls{io} & 0.5590 & 0.04110 & 0.0015 & 4.5863 \\
    \bottomrule
  \end{tabular}
\end{table}

\begin{figure}[ht]
  \centering
  \subfloat[active \gls{fl} curve]{\includegraphics[width=0.5\textwidth,
    keepaspectratio]{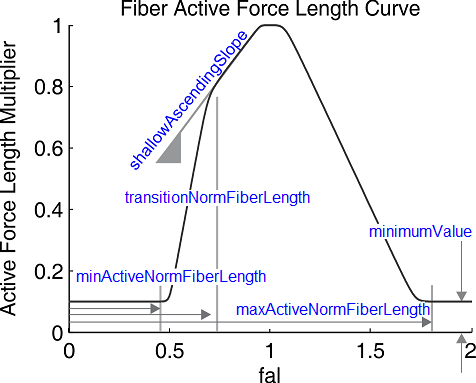}\label{fig:active-force-length-curve}}
  \subfloat[\gls{fl} curve]{\includegraphics[width=0.5\textwidth,
    keepaspectratio]{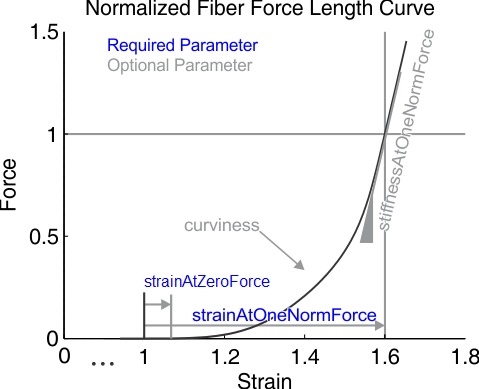}\label{fig:passive-force-length-curve}}
  \caption{The active \gls{fl} and \gls{fpe} curve definition for the Millard
    muscle model as implemented in \texttt{OpenSim}.}\label{fig:millard-curves}
\end{figure}

The active \gls{fl} and \gls{fpe} characteristic curves of the \gls{em} differ
significantly from those of a skeletal muscle. As shown in
\autoref{fig:millard-curves}, we can fine-tune the curve parameters so as to fit
the experimental data available for the \gls{lr} muscle. The values for the
active \gls{fl} and \gls{fpe} characteristic curves are summarized in
Tables~\ref{tab:eye-fl-curve} and~\ref{tab:eye-fpe-curve}, respectively. We
safely assume that the parameters of the characteristic curves for the other
\gls{em} are the same.

\begin{table}
  \centering
  \parbox{.45\linewidth}{
    \centering
    \caption{Parameters of the active \gls{fl} characteristic curve for the
      \gls{em}.}\label{tab:eye-fl-curve}
    \begin{tabular}{@{}cccccccccc@{}}
      \toprule
      \textbf{Parameter} & \textbf{Value} \\
      \midrule
      min norm active fiber length & 0.55 \\
      transition norm fiver length & 0.7 \\
      max norm active fiver length & 1.8 \\
      shallow ascending slope &  2.4 \\
      minimum value & 0.0 \\
      \bottomrule
    \end{tabular}
  }
  \quad
  \parbox{.45\linewidth}{
    \vspace{-1.5cm}
    \centering
    \caption{Parameters of the \gls{fpe} characteristic curve for the
      \gls{em}.}\label{tab:eye-fpe-curve}
    \begin{tabular}{@{}cccccccccc@{}}
      \toprule
      \textbf{Parameter} & \textbf{Value} \\
      \midrule
      strain at zero force & -0.18 \\
      strain at one norm force & 0.4 \\
      \bottomrule
    \end{tabular}
  }
\end{table}

\gls{em} have a higher fraction of fast twitch fibers and thus different
\gls{fv} behavior, due to different structures compared to skeletal
muscles. Despite that, the default Millard \gls{fv} curve was used for the six
\gls{em}, since the behavior of the selected muscle model depends heavily on the
maximum contraction velocity $v^{\text{max}}$. The maximum muscle contraction
velocity is tuned so as to match the peak velocity of saccadic eye movement
$\omega^{\text{max}} = 15.7 \si{\radian / \s}$ ($900 \si{\degree /
  \s}$). Following this definition, the maximum muscle contraction velocity is
given in optimal fiber length per seconds and it is thus different for each
\gls{em}, as their optimal fiber length is different
($v^{\text{max}} = \omega^{\text{max}} r / l_o^M$). Furthermore, since the optic
nerve is much shorter that the average muscle nerve, activation and deactivation
delays ($\tau_a = \tau_d = 5 \si{\milli\s}$) are smaller. Finally, two
separate wrapping spheres for the rectus muscles and the oblique muscles were
created, to avoid abnormal changes on the \gls{fl} curve as the eye rotates.

\subsection*{Passive Connective Tissues}\label{sec:passive-connective-tissues}

The passive connective tissues of the orbit apply a restoring force, which
brings the eye back to the central position when the net force from the \gls{em}
is zero. These tissues include all non-muscular suspensory tissues, such as
Tenon's capsule, the optic nerve, the fat pad and the conjunctiva. The
force-displacement elasticity force can be represented as

\begin{equation}\label{equ:passive-tissue}
  \vec{f}_t = -k_p \vec{q} - k_c \vec{q}^3 - k_d \vec{\dot{q}}
\end{equation}
where, $\vec{f}_t$ is the passive tissue forces,
$k_p= 2.225 \cdot 10^{-3} \si{\N \m / \radian}$,
$k_c= 34.53 \cdot 10^{-3} \si{\N \m / \radian^3}$ and
$k_v= 2 \cdot 10^{-3} \si{\N \m / (\radian / \s)}$ the physiological
constants~\cite{Robinson1969a, Collins1981, Priamikov2016} and
$\vec{q}, \vec{\dot{q}} \inr{3}$ the rotational coordinates and velocities of
the mode. These forces are modeled using \texttt{OpenSim}'s expression based
coordinate force.


\section*{Results}\label{sec:results}

\subsection*{Fixation Controller}\label{sec:fixation-controller}

A fixation controller that calculates the \gls{em} excitations required to track
a desired saccade was implemented. The controller actuates the model in a
closed-loop \gls{fd} manner. The parameters of the controller are the desired
horizontal and vertical fixation angles, the saccade onset and velocity, and the
gains of \gls{pd} tracking controller. A sigmoid function is used for generating
smooth saccade trajectories in the horizontal and vertical direction, while the
torsional component is maintained close to zero. More formally,

\begin{equation}\label{equ:sigmoid}
  \begin{aligned}
    \theta_d(t) &= \frac{a}{2} \Big(\tanh(b (t - t_0)) + 1\Big) \\
    \dot{\theta}_d(t) &= \frac{a b}{2} \Big(1 - \tanh^2(b (t - t_0))\Big)
  \end{aligned}
\end{equation}
where $\theta_d(t)$ and $\dot{\theta}_d(t)$ represent the desired orientation
and velocity at time $t$, $a$ the magnitude of the trajectory, $b$ the slope and
$t_0$ a time shift constant. Provided a fixation goal $\theta_g$, a desired
saccade velocity $\dot{\theta}_g$ and a saccade onset $t_g$, the parameters of
the sigmoid function are defined as $a = \theta_g$,
$b = 2 \dot{\theta}_g / \theta_g$ and $t_0 = t_s$. The output of the \gls{pd}
tracking controller has the following form

\begin{equation}\label{equ:eye-pd-controller}
  u(t) = k_p (\theta_d(t) - \theta(t)) + k_d (\dot{\theta}_d(t) -
  \dot{\theta}(t))
\end{equation}
where $k_p$, $k_d $ are the tracking gains, and $\theta(t)$, $\dot{\theta}(t)$
the simulated response of the model.

\begin{figure}[ht]
  \centering
  \subfloat{\includegraphics[height=.25\textheight]{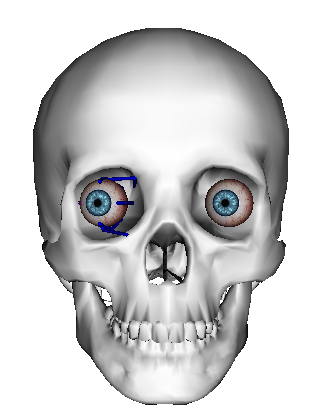}}
  \hspace{3cm}
  \subfloat{\includegraphics[height=.25\textheight]{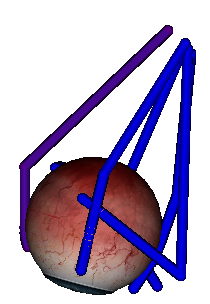}}
  \caption{Model with a fixation target at $\theta_H = -15 \si{\degree}$,
    $\theta_V = 0 \si{\degree}$ during simulation. Blue denotes low and red high
    muscle activation levels.}\label{fig:model}
\end{figure}

\begin{figure}[ht]
  \centering
  \subfloat[$\theta_H = -15 \si{\degree}$, $\theta_V = 0 \si{\degree}$, $b = 600
  \si{\degree / \s}$ and $k_v = 0.002 \si{\N \m / (\radian / \s)}$]
  {\includegraphics[trim={3cm 0cm 2.8cm 0cm},clip,width=1.\textwidth]{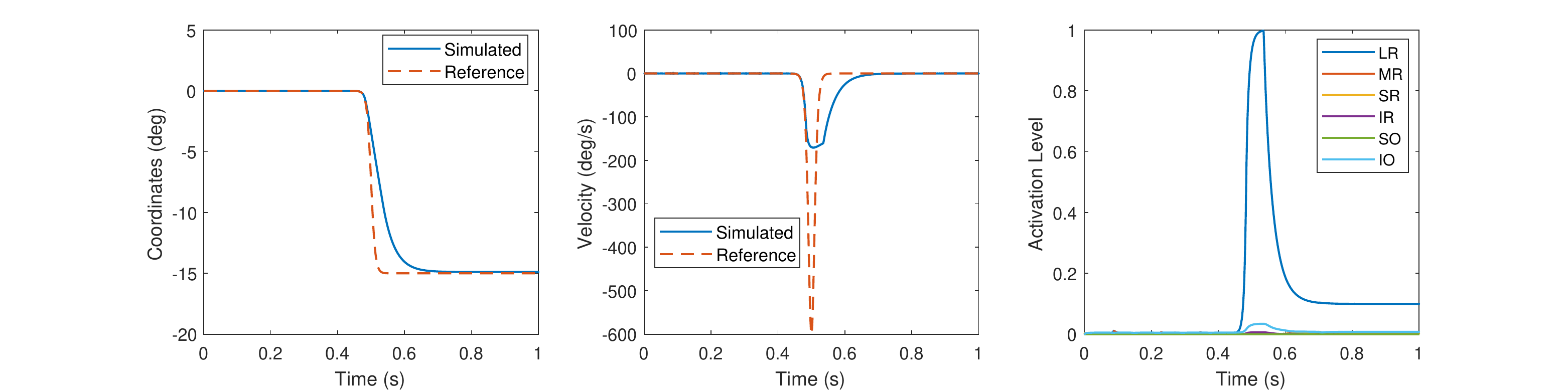}}  \\
  \subfloat[$\theta_H = -15 \si{\degree}$, $\theta_V = 0 \si{\degree}$, $b = 600
  \si{\degree / \s}$ and $k_v = 0 \si{\N \m / (\radian / \s)}$]
  {\includegraphics[trim={3cm 0cm 2.8cm 0cm},clip,width=1.\textwidth]{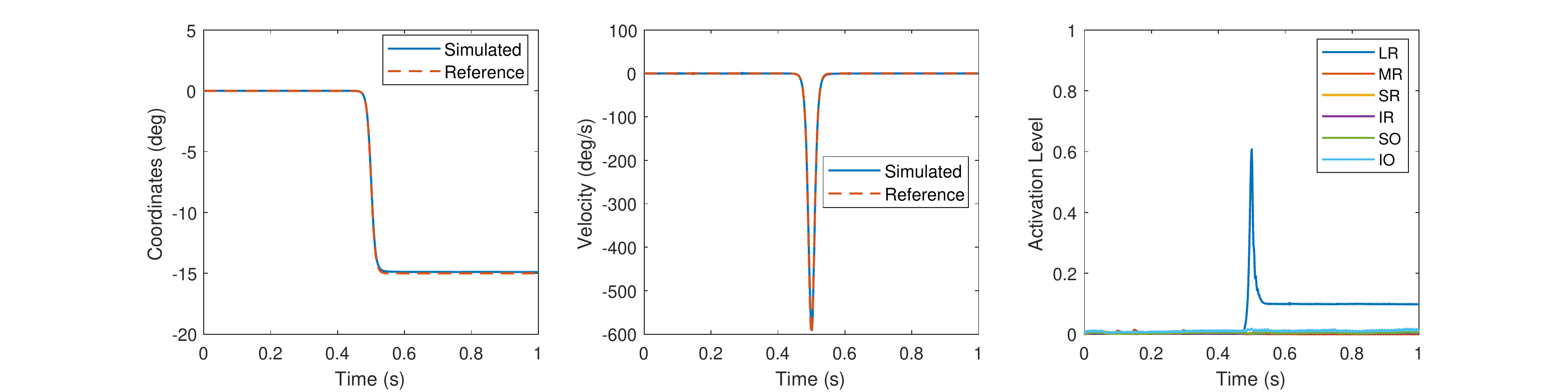}} 
  \caption{Simulated saccade response for different model parameters. Left
    subplot represents the simulated generalized coordinates, middle the
    generalized velocities and right the estimated \gls{em} excitation
    levels.}\label{fig:simulated-hor-saccade}
\end{figure}

\begin{figure}[ht]
  \centering
  \includegraphics[trim={3cm 0cm 2.8cm 0cm},clip,width=1.\textwidth]{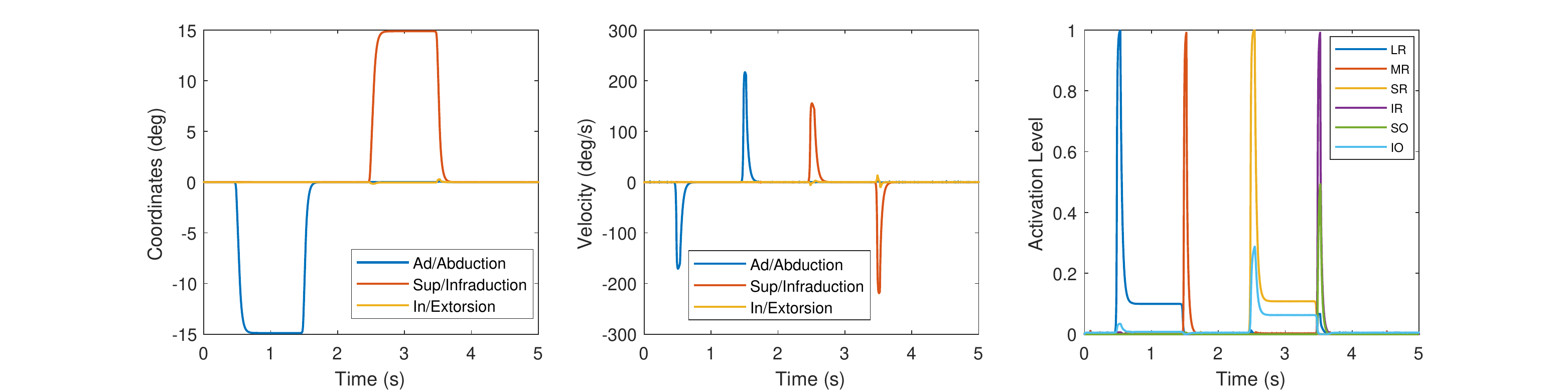}
  \caption{Simulated saccade response performing abduction/adduction
    ($0-2 \si{\s}$) and supraduction/infraduction ($2-4 \si{\s}$) with
    $k_v = 0.002 \si{\N \m / (\radian /
      \s)}$.}\label{fig:simulated-cmplx-saccade}
\end{figure}

The sign and magnitude of $u(t)$, representing the deviation from
the fixation target for each axis of rotation respectively, are used to
calculate the muscle excitation levels, by assuming that each individual muscle
rotates the eye globe in a particular direction. \autoref{fig:model} presents an
instance of the model during simulation with the corresponding muscles
activated. \autoref{fig:simulated-hor-saccade} depicts the simulated
coordinates, angular velocities and estimated \gls{em} excitation levels that
reproduce the desired saccade trajectory for different model
parameters. Finally, \autoref{fig:simulated-cmplx-saccade} shows alternations in
the saccadic movements both in the horizontal and vertical direction so as to
examine the activation and deactivation patterns of the \gls{em}.

\section*{Conclusion}\label{sec:concluison}

A realistic oculomotor model representing the motility of a normal human eye was
presented and made publicly available. The parameters of the model were
calibrated using available experimental measured data. The model can be used for
kinematics and dynamics analysis or as a tool for obtaining the muscle
activations that generate a desired saccade, using a closed-loop fixation
controller in a \gls{fd} manner. There is of course space for further
improvement, which will enhance the accuracy and the predictability of the
proposed computational model. In this study, we didn't attempt to model the
muscle pulleys~\cite{Kono2002a}, where the position of the pulleys vary as a
function of the model coordinates. Therefore, the users should consider
performing further validation of the eye model based on the requirements of the
targeted utility and the variables of interests.


\bibliography{mylibrary}

\begin{thebibliography}{18}
\providecommand{\natexlab}[1]{#1}
\providecommand{\url}[1]{#1}
\csname url@samestyle\endcsname
\providecommand{\newblock}{\relax}
\providecommand{\bibinfo}[2]{#2}
\providecommand{\BIBentrySTDinterwordspacing}{\spaceskip=0pt\relax}
\providecommand{\BIBentryALTinterwordstretchfactor}{4}
\providecommand{\BIBentryALTinterwordspacing}{\spaceskip=\fontdimen2\font plus
\BIBentryALTinterwordstretchfactor\fontdimen3\font minus
  \fontdimen4\font\relax}
\providecommand{\BIBforeignlanguage}[2]{{%
\expandafter\ifx\csname l@#1\endcsname\relax
\typeout{** WARNING: IEEEtranN.bst: No hyphenation pattern has been}%
\typeout{** loaded for the language `#1'. Using the pattern for}%
\typeout{** the default language instead.}%
\else
\language=\csname l@#1\endcsname
\fi
#2}}
\providecommand{\BIBdecl}{\relax}
\BIBdecl

\bibitem[Lee and Terzopoulos(2006)]{Lee2006}
S.-h. Lee and D.~Terzopoulos, ``{Heads Up ! Biomechanical Modeling and
  Neuromuscular Control of the Neck},'' \emph{ACM Transactions on Graphics},
  vol.~1, no. 212, pp. 1188--1198, 2006.

\bibitem[Delp et~al.(2007)Delp, Anderson, Arnold, Loan, Habib, John,
  Guendelman, and Thelen]{Delp2007}
S.~L. Delp, F.~C. Anderson, A.~S. Arnold, P.~L. Loan, A.~Habib, C.~T. John,
  E.~Guendelman, and D.~G. Thelen, ``{OpenSim : Open-Source Software to Create
  and Analyze Dynamic Simulations of Movement},'' \emph{IEEE Transactions on
  Biomedical Engineering}, vol.~54, no.~11, pp. 1940--1950, 2007.

\bibitem[Robinson et~al.(1969)Robinson, O'meara, Scott, and
  Collins]{Robinson1969a}
D.~A. Robinson, D.~M. O'meara, A.~B. Scott, and C.~C. Collins, ``{Mechanical
  components of human eye movements},'' \emph{Journal of Applied Physiology},
  vol.~26, no.~5, pp. 548--553, 1969.

\bibitem[Collins et~al.(1981)Collins, Carlson, Scott, and
  Jampolsky]{Collins1981}
C.~C. Collins, M.~R. Carlson, a.~B. Scott, and a.~Jampolsky, ``{Extraocular
  muscle forces in normal human subjects.}'' \emph{Investigative Ophthalmology
  {\&} Visual Science}, vol.~20, no.~5, pp. 652--664, 1981.

\bibitem[Iskander et~al.(2018)Iskander, Hossny, Nahavandi, del Porto, and
  Porto]{Iskander2018}
J.~Iskander, M.~Hossny, S.~Nahavandi, L.~del Porto, and L.~Porto, ``{An ocular
  biomechanic model for dynamic simulation of different eye movements},''
  \emph{Journal of Biomechanics}, 2018.

\bibitem[Robinson(1964)]{Robinson1964a}
D.~A. Robinson, ``{The mechanics of human saccadic eye movement},'' \emph{The
  Journal of Physiology}, pp. 245--264, 1964.

\bibitem[Robinson and Fuchs(1969)]{Robinson1969}
D.~A. Robinson and A.~F. Fuchs, ``{Eye movements evoked by stimulation of
  frontal eye fields.}'' \emph{Journal of Neurophysiology}, vol.~32, no.~5, pp.
  637--648, 1969.

\bibitem[Thelen et~al.(2003)Thelen, Anderson, and Delp]{Thelen2003a}
D.~G. Thelen, F.~C. Anderson, and S.~L. Delp, ``{Generating dynamic simulations
  of movement using computed muscle control},'' \emph{Journal of Biomechanics},
  vol.~36, no.~3, pp. 321--328, mar 2003.

\bibitem[Millard et~al.(2013)Millard, Uchida, Seth, and Delp]{Millard2013}
M.~Millard, T.~Uchida, A.~Seth, and S.~L. Delp, ``{Flexing computational
  muscle: modeling and simulation of musculotendon dynamics},'' \emph{Journal
  of Biomechanical Engineering}, vol. 135, no.~2, pp. 1--12, mar 2013.

\bibitem[James et~al.(2018)James, Papapavlou, Blenkinsop, Cope, Anderson,
  Moustakas, and Gurney]{James2018}
S.~S. James, C.~Papapavlou, A.~Blenkinsop, A.~J. Cope, S.~R. Anderson,
  K.~Moustakas, and K.~N. Gurney, ``{Integrating brain and biomechanical
  models-A new paradigm for understanding neuro-muscular control},''
  \emph{Frontiers in Neuroscience}, vol.~12, no. FEB, 2018.

\bibitem[Priamikov et~al.(2016)Priamikov, Fronius, Shi, and
  Triesch]{Priamikov2016}
A.~Priamikov, M.~Fronius, B.~Shi, and J.~Triesch, ``{OpenEyeSim: A
  biomechanical model for simulation of closed-loop visual perception},''
  \emph{Journal of Vision}, vol.~16, no.~15, p.~25, dec 2016.

\bibitem[Kono et~al.(2002)Kono, Clark, and Demer]{Kono2002a}
R.~Kono, R.~A. Clark, and J.~L. Demer, ``{Active pulleys: Magnetic resonance
  imaging of rectus muscle paths in tertiary gazes},'' \emph{Investigative
  Ophthalmology and Visual Science}, vol.~43, no.~7, pp. 2179--2188, 2002.

\bibitem[Clark et~al.(1977)Clark, Miller, and Demer]{Clark1977}
R.~A. Clark, J.~M. Miller, and J.~L. Demer, ``{Three-dimensional Location of
  Human Rectus Pulleys by Path Inflections in Secondary Gaze Positions},''
  \emph{Investigative Ophthalmology {\&} Visual Science}, vol.~41, no.~12, pp.
  3787--3797, nov 1977.

\bibitem[Miller(2007)]{Miller2007}
J.~M. Miller, ``Understanding and misunderstanding extraocular muscle
  pulleys,'' \emph{Journal of Vision}, vol.~7, no.~11, p.~10, 2007.

\bibitem[Wei et~al.(2010)Wei, Sueda, and Pai]{Wei2010a}
Q.~Wei, S.~Sueda, and D.~K. Pai, ``{Physically-based modeling and simulation of
  extraocular muscles},'' \emph{Progress in Biophysics and Molecular Biology},
  vol. 103, no. 2-3, pp. 273--283, dec 2010.

\bibitem[Bach-y Rita et~al.(1971)Bach-y Rita, Collins, {Smith-Kettlewell
  Institute of Visual Sciences.}, and {University of the Pacific. Department of
  Visual Sciences.}]{Bach-y-Rita1971}
P.~Bach-y Rita, C.~C. Collins, {Smith-Kettlewell Institute of Visual
  Sciences.}, and {University of the Pacific. Department of Visual Sciences.},
  \emph{{The control of eye movements.}}\hskip 1em plus 0.5em minus 0.4em\relax
  Academic Press, 1971.

\bibitem[Angelaki and Hess(2004)]{Angelaki2004}
D.~E. Angelaki and B.~J.~M. Hess, ``{Control of eye orientation : where does
  the brain ' s role end and the muscle ' s begin ?}'' \emph{Neuroscience},
  vol.~19, 2004.

\bibitem[Wong(2004)]{Wong2004}
A.~M. Wong, ``{Listing's law: clinical significance and implications for neural
  control},'' \emph{Survey of Ophthalmology}, vol.~49, no.~6, pp. 563--575, nov
  2004.

\end{thebibliography}

\end{document}